\documentclass[superscriptaddress,amsmath,amssymb,aps,prd,section,twocolumn]{revtex4-2}

\newcounter{bla}

\usepackage{graphicx}
\usepackage{bm}
\usepackage{amssymb}
\usepackage{amsmath}
\usepackage{amsfonts}
\usepackage[utf8]{inputenc}
\usepackage[colorlinks=true,allcolors=blue]{hyperref}
\usepackage[capitalise]{cleveref}

\usepackage{siunitx}
\usepackage{xspace}
\usepackage{comment}
\usepackage{float}

\usepackage{chngcntr}

\newcommand{\abs}[1]{\left| #1 \right|}

\DeclareFixedFont{\ttb}{T1}{txtt}{bx}{n}{12} 
\DeclareFixedFont{\ttm}{T1}{txtt}{m}{n}{12}  

\usepackage{color}
\definecolor{deepblue}{rgb}{0,0,0.5}
\definecolor{deepred}{rgb}{0.6,0,0}
\definecolor{deepgreen}{rgb}{0,0.5,0}

\begin{document}
\title{Cascade Appearance Signatures of Sterile Neutrinos at 1-100 TeV}
\author{B. R. Smithers}
\thanks{Corresponding author: \url{benjamin.smithers@mavs.uta.edu}}
\affiliation{Department of Physics, University of Texas at Arlington,
Arlington, Texas 76019, USA}
\author{B. J. P. Jones}
\affiliation{Department of Physics, University of Texas at Arlington,
Arlington, Texas 76019, USA}
\author{C. A. Arg\"uelles}
\affiliation{Department of Physics, Harvard University, Cambridge, Massachusetts, USA}
\author{J. M. Conrad}
\affiliation{Dept. of Physics, Massachusetts Institute of Technology, Cambridge, MA 02139, USA}
\author{A. Diaz}
\affiliation{Dept. of Physics, Massachusetts Institute of Technology, Cambridge, MA 02139, USA}

\date{\today}

\begin{abstract}
Neutrino telescopes provide strong sensitivity to sterile neutrino oscillations through matter-effects occurring in the few TeV energy range for eV$^{2}$-scale neutrino mass-squared splittings.  
Prior searches have focused on $\nu_\mu$ disappearance, which has a particularly strong sensitivity to the mixing angle $\theta_{24}$ via $\nu_\mu\rightarrow\nu_s$ transitions.  
Nowadays, the $\nu_\mu\rightarrow\nu_e$ and $\nu_\mu\rightarrow\nu_\tau$ appearance channels have been considered less promising at neutrino telescopes, due in part to the much smaller target volume for cascades relative to tracks. 
This work explores the detectability of these signatures at neutrino telescopes given present constraints on sterile neutrino mixing, and as an example, forecasts the sensitivity of the IceCube Neutrino Observatory to the mixing angles $\theta_{14}$, $\theta_{24}$, and $\theta_{34}$ in the 3+1 sterile neutrino model using the cascade channel with ten years of data. 
We find  that $\nu_\tau$ appearance signatures consistent with the existing IceCube $\nu_\mu$ disappearance best-fit point are discoverable for values of $\theta_{34}$ consistent with world constraints, and that the sterile neutrino parameters favored by the BEST and gallium anomalies are expected to be testable at the 95\% confidence level. 
\end{abstract}

\maketitle

\section{\label{sec:one}Introduction}


The three-mass and three active-flavor neutrino paradigm has been well-studied~\cite{PhysRevD.98.030001,Esteban_2019,de_Salas_2018,Capozzi_2016,zboson2006, berns2021recent}.
However, several anomalies persist at short baselines, including in $\nu_\mu\rightarrow\nu_e $ appearance in decay-in-flight~\cite{aguilar2018significant} and decay-at-rest~\cite{Athanassopoulos_1998} beams  and $\nu_e\rightarrow\nu_e$ disappearance at reactors~\cite{mention2011reactor,serebrov2019first}  and with $^{71}$Ga electron capture sources~\cite{PhysRevC.73.045805,giunti2011statistical}.  These anomalies have been attributed to possible oscillations of unknown neutrinos with mass-squared differences in the range of $\Delta m^{2}\sim 0.1-10\text{ eV}^{2}$~\cite{abazajian2012light}.   Such an additional neutrino flavor state must be non-weakly interacting, or ``sterile,'' to be consistent with observed decay widths of the Z-boson~\cite{zboson2006}; the simplest such model is known as the ``3+1'' light sterile neutrino model in which a single sterile neutrino is added. 

There have been interesting recent developments for the 3+1 model.  The BEST experiment appears to validate the anomalous electron neutrino disappearance signature of the previous gallium anomalies with a new level of statistical significance and experimental precision~\cite{barinov2021results}. The Neutrino-4 experiment claims evidence of short-baseline oscillations in the $\bar{\nu}_e$ disappearance channel with $\Delta m^2\sim 7.3\,\mathrm{eV}^2$ at the 2.9$\sigma$ level. Meanwhile results from the MicroBooNE~\cite{microboonecollaboration2021search,microboonecollaboration2021search1,microboonecollaboration2021searchmulti} experiment challenge the interpretation that the MiniBooNE low energy excess~\cite{miniboone2018} is due entirely to the electron neutrino by placing a constraint on the sterile neutrino interpretation of the excess; though the impact of this observation on the 3+1 model is just beginning to be assessed~\cite{arguelles2021microboone,denton2021sterile}.  Continued exploration of sterile neutrino mixing in all channels and all energy ranges thus remains strongly motivated~\cite{sbnfermilab}.

The addition of a fourth neutrino mass and flavor eigenstate expands the unitary mixing matrix to four dimensions. The four-neutrino oscillations model becomes an extension of the three-neutrino model with three additional mixing angles $\theta_{14}$, $\theta_{24}$, and $\theta_{34}$, and two new CP-violating phases $\delta_{14}$ and $\delta_{24}$. These three new mixing angles parametrize the amplitude of oscillations between the three active states and the one sterile state, and lead to additional short-baseline vacuum-like oscillations as well as novel effects in the presence of matter~\cite{Akhmedov:1988kd,KRASTEV1989341,Chizhov:1998ug, Chizhov_1999, Akhmedov_2000}.  In this work we consider CP-conserving models with all CP-violating phases set to zero.

Of particular interest to neutrino telescopes, matter effects can result in the near complete disappearance of TeV-scale muon anti-neutrinos passing through the Earth's core for a sterile neutrino with eV-scale mass squared differences~\cite{Nunokawa:2003ep, Petcov:2016iiu, Choubey:2007ji, Barger:2011rc, Esmaili:2012nz, esmaili2013restricting, Lindner:2015iaa}. This signature of matter-enhanced resonant disappearance has been targeted by the IceCube Neutrino Observatory~\cite{Aartsen_2020, Aartsen_2020_prd}, leading to one of the  most sensitive $\nu_\mu$ disappearance analyses to date. The result of the analysis was a closed 90\% contour with best fit point at $\sin^2 2\theta_{24}\sim0.1$ and $\Delta m^2_{41}=4.5\text{ eV}^2$, under a conservative assumption (for the $\nu_\mu$ disappearance channel) that $\theta_{34}=\theta_{14}=0$. In addition to being a strong refutation, lower mass solutions consistent with the LSND~\cite{Athanassopoulos_1998} and MiniBooNE anomalies and constraints around 1~eV$^2$~\cite{kopp2013sterile, Cirelli:2004cz, abazajian2012light, Gariazzo:2017fdh, Dentler:2017tkw, Diaz:2019fwt}, a possible interpretation of this result is as a statistically weak hint of a disappearance signature around $\Delta m^2_{41}\sim4.5\text{ eV}^2$.  Further exploration of this region of parameter space  in other channels at neutrino telescopes is therefore strongly motivated. 

In this work, we explore the potential of sterile neutrino searches at gigaton-scale neutrino telescopes using matter-enhanced $\nu_\tau$ and $\nu_e$  appearance signatures that occur when either $\theta_{34}$ or $\theta_{14}$ are non-zero~\cite{esmaili2013}. We will show that $\nu_\tau$ appearance of considerable strength may accompany $\nu_\mu$ disappearance within the IceCube allowed region for $\Delta m^2_{41}$ and $\theta_{24}$, for values of $\theta_{34}$ that remain consistent with world data sets.  We will also demonstrate that these signatures can be probed using IceCube's public data samples.  Finally, we will also explore possible sensitivity to $\nu_e$ appearance at levels consistent with the gallium and BEST anomalies.

\begin{figure}
    \centering
    \includegraphics[width=0.95\linewidth]{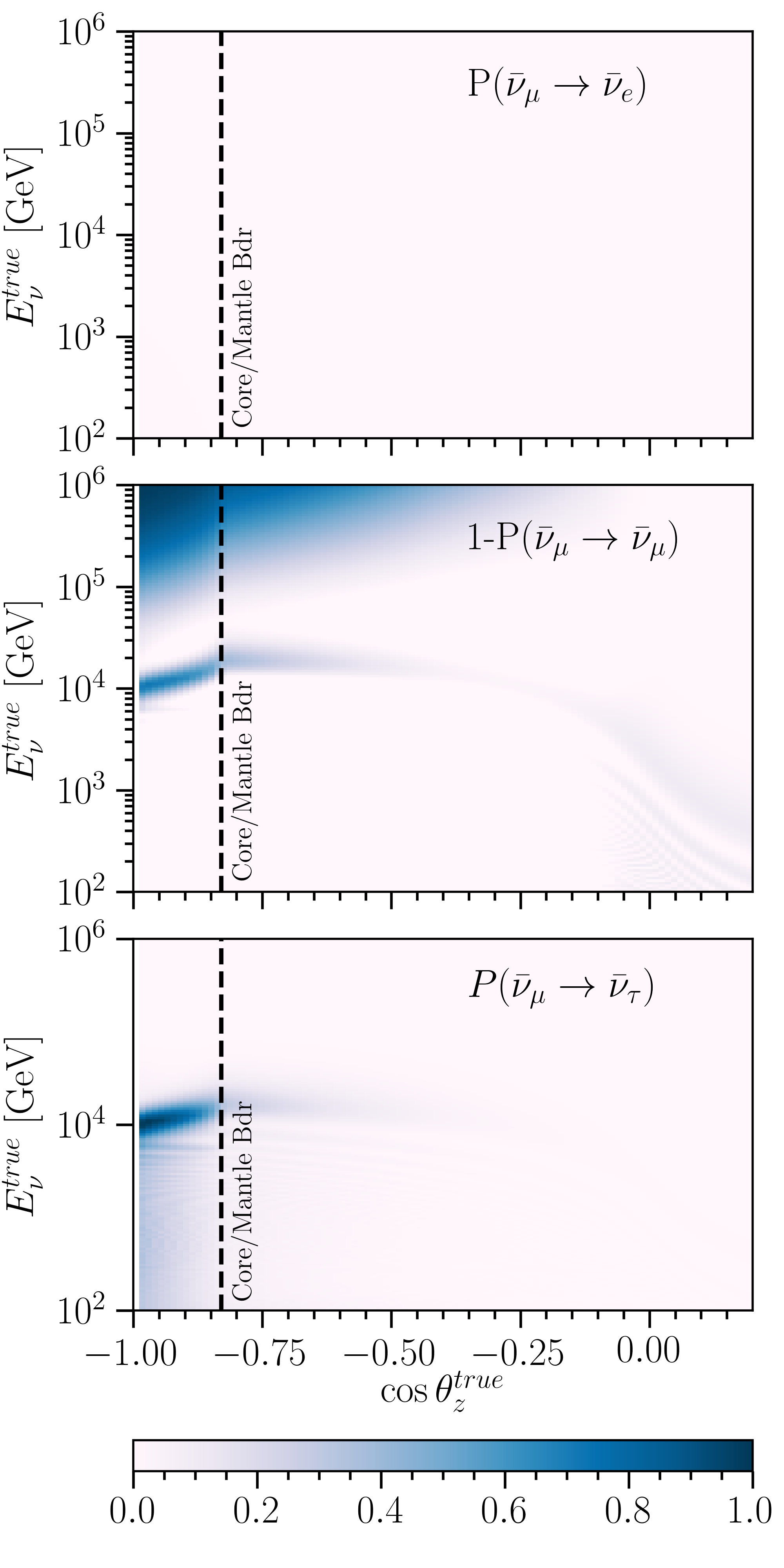}
    \caption{Transition probabilities $P(\bar{\nu}_{\mu}\to\bar{\nu}_{\alpha})$ for $\bar{\nu}_e$ (top), $\bar{\nu}_{\mu}$ (middle), and $\bar{\nu}_{\tau}$ (bottom) for a sterile neutrino flux with \(\sin^{2}(2\theta_{24})=0.1\), \(\sin^{2}(2\theta_{34})=0.2\), and \(\Delta m^{2}_{41}=4.5\text{ eV}^{2}\). A dashed black line is used to denote the outer core-mantle boundary, and a solid black line denotes the inner-outer core boundary. These probabilities are shown as a function of the neutrino's energy ($E_{\nu}^{true}$) and the cosine of the angle measured from an upwards direction, towards the neutrino's origin.}\label{fig:excess}
\end{figure}

The IceCube Neutrino Observatory is described at length in Ref.~\cite{Aartsen_2017}. Briefly, the detector is a cubic-kilometer Cherenkov neutrino observatory one and half kilometers deep in the Antarctic ice~\cite{Aartsen_2017}.
There, 5160 photo-multiplier tubes encased within glass pressure vessels, or ``Digital Optical Modules'' (DOMs)~\cite{ABBASI2009294} detect Cherenkov emission from charged particles traversing the ice.
The DOMs are arranged vertically with a seventeen meter spacing into seventy-nine strings, which themselves are aligned into a hexagonal lattice with a 125 meter spacing. 
An additional, more densely instrumented sub-detector called DeepCore exists towards the bottom-center of the main detector~\cite{ABBASI2012615}.
The observatory has been running for over a decade and has accumulated large numbers of $\nu_{\mu}$ CC interactions which make depositions of light that make long signatures in the detector called tracks; and neutral current, electron neutrino, and tau neutrino events which deposit light in blob-like shapes called cascades. These event topologies are elaborated upon in Section~\ref{sec:depo}.

\begin{figure}
    \centering
    \includegraphics[width=0.98\linewidth]{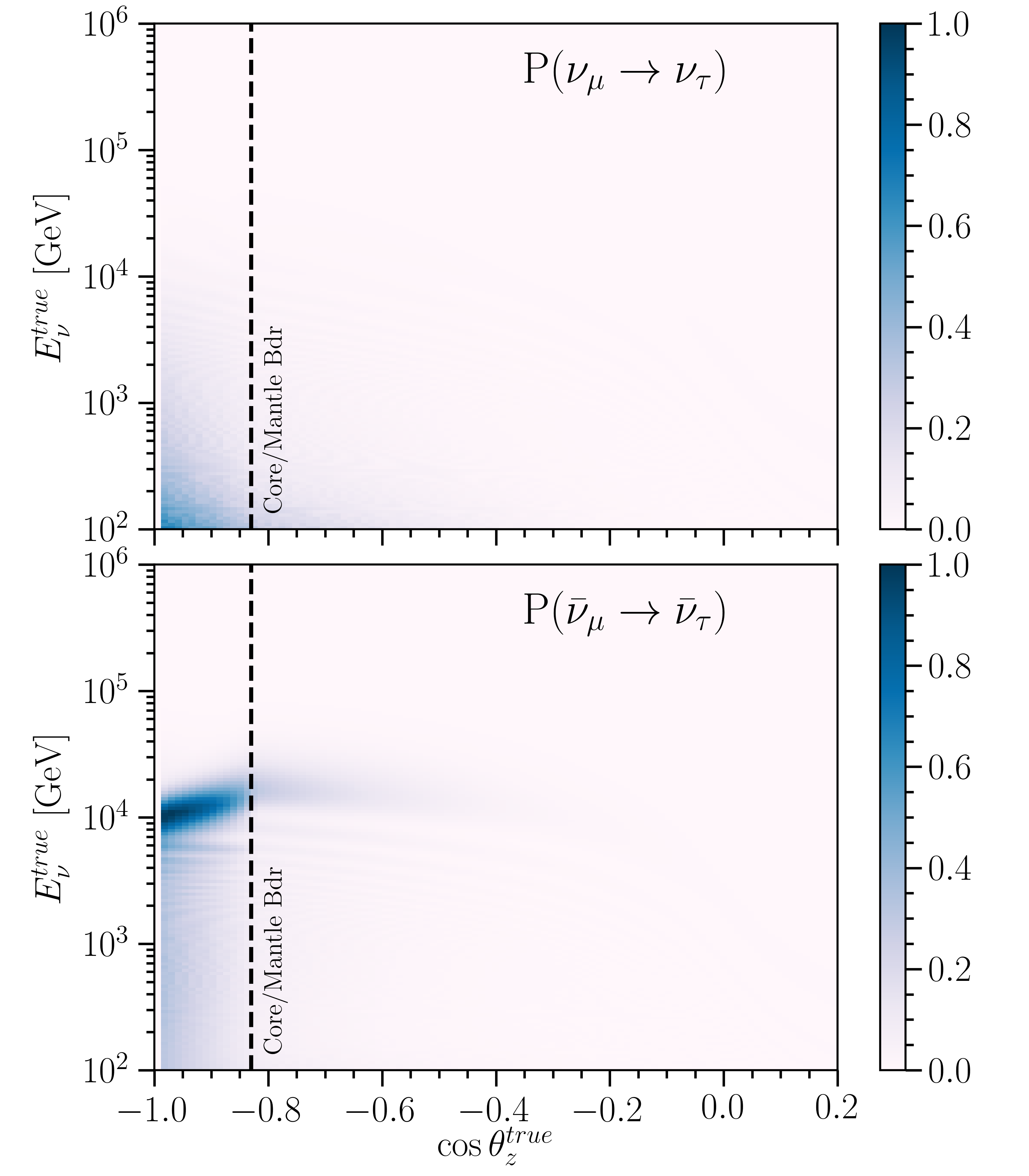}
    \caption{Appearance probabilities for $P(\nu_{\mu}\to\nu_{\tau})$ (top) and $P(\bar{\nu}_{\mu}\to\bar{\nu}_{\tau})$ (bottom) for a sterile neutrino flux with \(\sin^{2}(2\theta_{24})=0.1\), \(\sin^{2}(2\theta_{34})=0.19\), and \(\Delta m^{2}_{41}=4.5\text{ eV}^{2}\).}\label{fig:nunubar}
\end{figure}

IceCube analyses targeting $\nu_\mu$ disappearance are considered track-like only, since the only available signature under the previous mixing assumptions $\theta_{14}=\theta_{34}=0$ is $\nu_\mu\rightarrow\nu_s$ disappearance.  In similar models with both non-zero $\theta_{24}$ and $\theta_{34}$, however, resonant $\nu_{\mu}\rightarrow\nu_{\tau}$ oscillations lead to a strong appearance signature of $\nu_{\tau}$ as shown in Figure~\ref{fig:excess}. While some of the $\nu_{\tau}$ will produce $\tau^{\pm}$ that decay leptonically to produce additional tracks, dampening the $\nu_\mu$ disappearance signature, most charged current $\nu_\tau$ and $\bar{\nu}_\tau$ interactions will produce localized energy deposits that will be reconstructed as single cascades at these energies~\cite{abbasi2020measurement}. 
As in the $\nu_\mu\rightarrow\nu_s$ channel, the most striking feature of the signature is a resonant flavor oscillation for Earth-core-crossing anti-neutrinos at a specific energy, proportional to the sterile neutrino $\Delta m^2_{41}$ value.  Since this matter effect occurs because of an interference between the vacuum oscillation phase and the matter-driven phase, the latter changing sign between neutrinos and anti-neutrinos, for small mixing angles the resonance is only present in for anti-neutrinos, given a heavier sterile neutrino.  The appearance probabilities for $\nu_\mu\rightarrow\nu_{\tau}$ and $\bar{\nu}_\mu\rightarrow\bar{\nu}_{\tau}$ are shown separately in Figure~\ref{fig:nunubar}. 

For zero $\theta_{24}$ very little signal is expected since the muon neutrinos, which dominate the flux at IceCube, cease to mix with the heavier mass state. As a consequence there will be negligible $\nu_\tau$ appearance, regardless of the value of $\theta_{34}$.  However, recent IceCube results favor a non-zero value for $\sin^2(2\theta_{24})$ of around 0.1, and assuming $\nu_\mu / \nu_4$ mixing at this level, the observable $\nu_\tau$ appearance will depend strongly on the value of $\theta_{34}$.  At the smallest values of $\theta_{34}$ ($\theta_{34}\lesssim 0.1 $), $\nu_{\mu}\to\nu_{s}$ oscillations dominate over $\nu_{\mu}\to\nu_{\tau}$ appearance from standard oscillations, and $\nu_\mu$ disappearance is the only visible signature. For values of $\theta_{34}$ larger than this threshold, the $\nu_{\mu}\to\nu_{\tau}$ oscillations begin to dominate and $\nu_{\tau}$ appearance manifests, leading to the appearance signature shown in Figure~\ref{fig:excess} (bottom). Increasing $\Delta m_{41}^{2}$ has the effect of broadening the appearance signature until $\sim 10\text{ eV}^{2}$, after which raising the mass-squared splitting has only a marginal effect. Increasing $\theta_{24}$ while reducing $\theta_{34}$ proportionately leaves the appearance signatures mostly unchanged while diminishing the disappearance amplitude.  

In addition, the effects of non-zero mixing angle $\theta_{14}$ can also be considered, having the consequence of introducing similar appearance signatures into the $\nu_e$ appearance channel~\cite{wang2021search}. For practical purposes $\nu_e$ and $\nu_\tau$ charged current events are indistinguishable at these energies in IceCube.   Notably, neutrino telescopes are the only experiments in the world with substantial sensitivity to sterile-neutrino induced $\nu_\tau$ appearance, so such direct constraints on the $\theta_{34}$ parameter are specific to these programs. Constraints on $\theta_{14}$, on the other hand, may directly relate to anomalies in $\nu_\mu\rightarrow\nu_e$ and $\nu_e$ disappearance. As we will show, constraining the $\nu_e$ appearance signature at IceCube under the non-zero presently favored value of $\theta_{24}$ from $\nu_\mu$ disappearance have direct implications for the BEST anomaly and the associated reactor-$\bar{\nu}$ anomaly.

This rich phenomenology motivates multi-mixing-angle and multi-channel searches to fully explore sterile neutrino mixing around the matter resonance at neutrino telescopes. In this work, we explore this space of mixing parameters by using publicly available tools, effective areas, and Monte Carlo simulation to estimate IceCube's sensitivity to $\theta_{24}$, $\theta_{34}$, and $\Delta m_{41}^{2}$ through cascades.

\section{\label{sec:depo} Neutrino Energy Deposition}

Large-volume neutrino telescopes typically are sensitive in the TeV to PeV energies; here, Deep-Inelastic Scattering (DIS)~\cite{gandhineutrinos} and the recently-observed~\cite{IceCube:2021rpz} Glashow-Resonance~\cite{PhysRev.118.316} interactions dominate. 
The detected neutrino interaction events fall into two morphological categories: tracks and cascades.

Charged-current (CC) $\nu_{\mu}$ DIS events result in muons at energies where radiative processes dominate energy loss rates.
As a result, energy losses are stochastically driven and the produced muons travel for kilometers. 
The results are threefold: muons are difficult to fully contain in neutrino telescopes, muon energies are poorly correlated with progenitor muon-neutrino energies, and muons' long travel-distance can allow for reconstructing their direction to within 1$^{\circ}$~\cite{trackaccuracy2017}. These events are called \textit{tracks}~\cite{icecube_energy_reco}.

All neutral-current (NC) DIS events result in a hadronic shower spreading around the interaction point and a secondary neutrino invisibly carrying away a proportion of the parent neutrino's energy. 
These events are often contained with a spherical topology. 
$\nu_{e}$-CC interactions develop similarly to neutral-current interactions, but repeated inverse Compton scattering of the produced electron initiates an electromagnetic shower superimposed over the hadronic shower. 
Thus, nearly all of the interacting neutrino's energy is observable as detectable light. 
These events are called \textit{cascades.} Such events tend to be well-contained permitting an efficient energy reconstruction, although suffer from poor angular reconstruction~\cite{icecube_energy_reco}. 

The evolution of a $\nu_{\tau}$-CC interaction is highly dependent on the energies involved. A tau is produced simultaneously as a hadronic cascade propagates around the interaction point, and then the tau decays. 
Due to their large mass, taus have a short lifetime and a decay length of $\sim 50$ m per PeV of tau energy~\cite{abbasi2020measurement}. 
From the tau branching ratios~\cite{PhysRevD.98.030001}, 17.37\% of the charged tau decays evolve as muon tracks, while the remainder of the decays evolve as electromagnetic or hadronic cascades. Only at neutrino energies above 60 TeV do $\nu_{\tau}$-CC interactions yield events with distinguishable primary and secondary cascades~\cite{abbasi2020measurement}.

Several distinct event samples have been developed to study these different types of events in IceCube. The High-Energy Starting Events sample~\cite{2021hese}, for example, was developed to study both taus and high-energy neutrinos likely astrophysical in origin. There exist other events samples optimized for higher event rates at lower energies, such as the Medium-Energy Starting Events~\cite{PhysRevDoverone}, and the five-year inelasticity sample~\cite{inelasticity2019}. There are also samples optimized for muon purity, such as the eight-year atmospheric muon sample~\cite{Aartsen_2020_prd} and others optimized for accurate energy resolution such as the six-years cascade sample~\cite{sixyrscascade}. This work will consider the cascade event selection described in~\cite{2018PhDT17N} and the track event selection previously used in IceCube sterile neutrino searches~\cite{PhysRevLett.117.071801}.

\section{\label{sec:flux} Neutrino Fluxes}

We calculate the expected event rates in IceCube exclusively using publicly available data on effective areas and publicly available Monte Carlo simulation samples. By studying the expected event rates in both track and cascade channels, we are able to estimate IceCube's sensitivities to sterile neutrino parameters given the existing ten year data set. 
At sensitive energies there are two relevant neutrino populations whose flux must be modeled: atmospheric and astrophysical neutrinos. 

Predicting atmospheric neutrino event rates requires a progenitor cosmic-ray flux, simulation of resulting air showers, propagation of the shower-born neutrinos through the Earth, and convolution of these fluxes with effective areas for a given sample selection to yield a final predicted event rate. 
For this work, we use the MCEq cascade equation solver~\cite{fedynitch2015calculation} with the three-population Hillas-Gaisser 2012 H3a cosmic-ray flux model~\cite{GAISSER2012801} and using the SYBILL 2.3c hadronic interaction model~\cite{Riehn:2017mfm} to simulate air showers.
The Poly-Gonato model for the cosmic-ray flux~\cite{polygonato2003} and QGSJET-II-04 model for hadronic interactions~\cite{qgsjet2011} were also found to produce similar results for this analysis. 

These fluxes are then propagated through the Earth using the Simple Quantum Integro-Differential Solver for neutrino oscillations (nuSQuIDS)~\cite{Delgado:2014kpa,arguelles:2015nu,arguelles2021nusquids}. 
We have configured nuSQuIDS to propagate the fluxes according to a spherically-symmetric Preliminary Reference Earth Model (PREM)~\cite{DZIEWONSKI1981297}; where it accounts for both coherent and non-coherent interactions relevant at these energies~\cite{Gonzalez_Garcia_2005, PhysRev.118.316} as well as tau-neutrino regeneration~\cite{PhysRevLett.81.4305}. 
For this work we use the CSMS cross sections~\cite{CooperSarkar:2011pa}. We fix the three-neutrino oscillations parameters to their global best-fit values~\cite{nufit2020}.

Astrophysical neutrino event rates are calculated similarly. Although, the neutrino flux prior to propagation through the Earth instead is expected to follow a power-law spectrum as a function of neutrino energy $E_{\nu}$, 
\begin{equation}
\Phi_{astr,\alpha}(E_{\nu}) = r_{\alpha}\Phi_{0}\left(\dfrac{E_{\nu}}{E_{0}}\right)^{-\gamma},
\end{equation}
normalized at $E_{0}=$100 TeV and with $\Phi_{0}=2.85\times 10^{-18}[\text{GeV}\cdot\text{cm}^{2}\cdot\text{sr}\cdot\text{s}]^{-1}$, a spectral index of $\gamma=2.39$~\cite{Aartsen_2020_prd}, and a flavor-ratio $r_{\alpha}$ for $\alpha\in( e,\mu,\tau)$. The flux is assumed isotropic and to have a $\nu:\bar{\nu}:$ ratio of $1:1$. 
Astrophysical neutrinos are assumed to be created with regards to the pion-decay induced flavor ratio of 1:2:0~\cite{PhysRevD.68.093005,ATHAR_2006}; these are then propagated through vacuum over large energy-baseline ratios, recovering the expected $\tfrac{1}{3}:\tfrac{1}{3}:\tfrac{1}{3}$ flavor ratio at Earth for the three-neutrino model~\cite{Aartsen_2015}.  
The same is done for sterile-neutrino hypotheses to predict expected four-flavor flavor ratios~\cite{carlos2020}.

\begin{figure}
    \centering
    \includegraphics[width=0.95\linewidth]{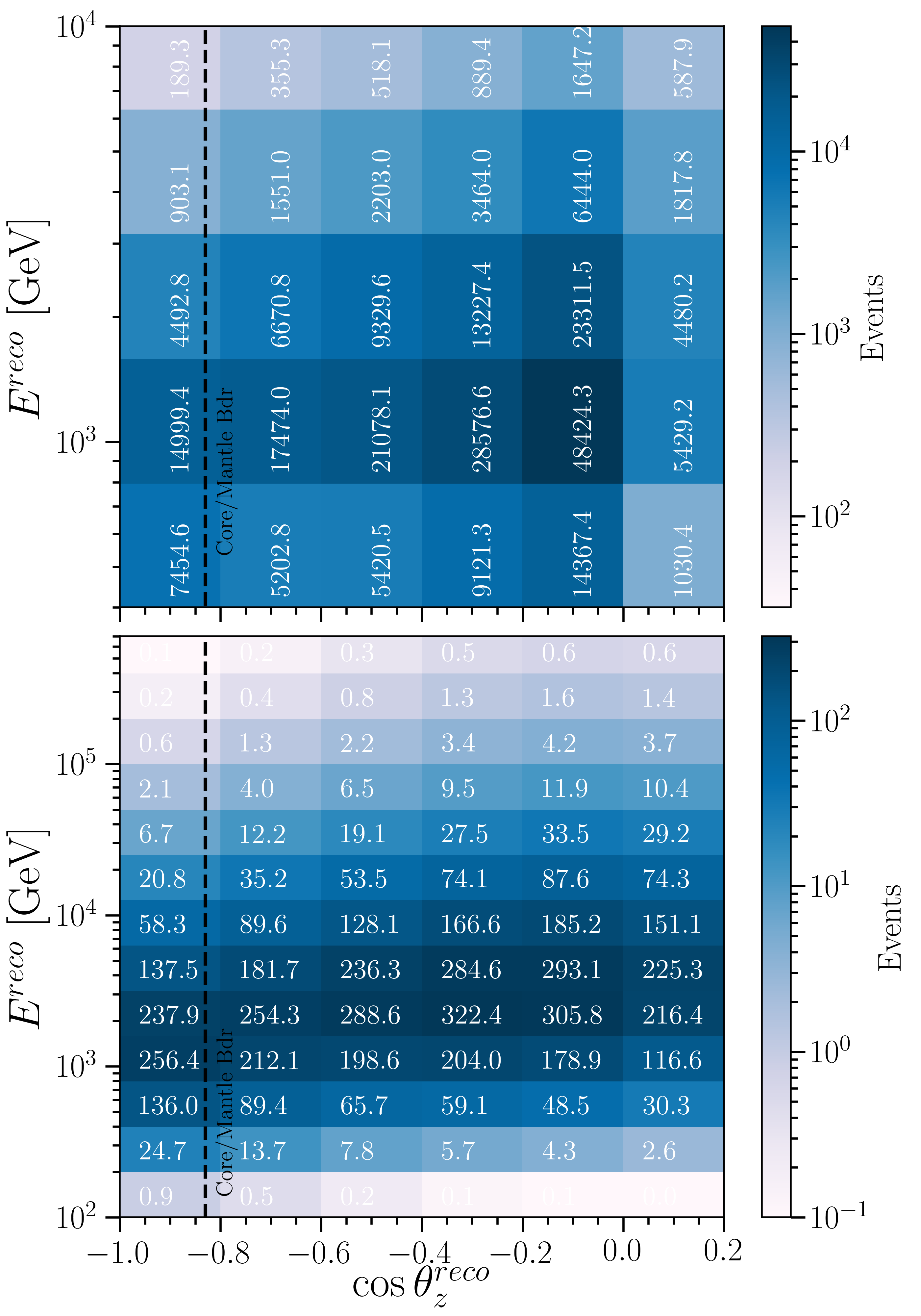}
    \caption{Expected number of through-going tracks (top) and cascades (bottom) for ten years of livetime using the Hillas Gaisser H3a cosmic ray flux model, SYBILL 2.3c interaction model, and the event selection described in Ref~\cite{2018PhDT17N}.}\label{fig:eventrate}
\end{figure}

\subsection{\label{sub:cascade} Cascade Rates}

Total cascade event rates in IceCube are calculated bin-wise, linearly in $\log(E^{\text{true}})$ and $\cos\theta_{z}^{\text{true}}$, by integrating over a product of flux and effective area $A_{eff}$ in each bin (i,j), summing for each neutrino species \(\alpha\), and multiplying by livetime $\tau$. This is shown below in Equation~\ref{eq:goodrate}. 
\begin{equation}\label{eq:goodrate}\begin{split}
N^{evt}_{i,j} = 2\pi\tau\sum_{\alpha}\int\limits_{E_{i}}^{E_{i+1}} & dE^{true}_{\nu}\int\limits_{\cos\theta_{j}}^{\cos\theta_{j+1}} d(\cos\theta_{z}^{true})\\
& \times\Phi_{\alpha}(E_{\nu}^{true},\cos\theta_{z}) A_{eff, \alpha}(E_{\nu}^{true}, \cos\theta_{z}^{true})
\end{split}\end{equation}
The effective areas used are publicly available and determined from the gradient boosted decision tree methods event selection developed and available 
in Ref.~\cite{2018PhDT17N}.
Expected bin-wise event counts $N^{mn}$ at reconstructed energy $(E_{\nu}^{reco})_{m}$ and zenith $(\cos\theta_{z}^{reco})_{n}$ follow from smearing from these expected true values by a bin-to-bin reconstruction probability $P_{mn}^{ij}$,
\begin{equation}
N_{mn}^{reco}= N_{ij}^{true} P_{ijmn}.
\end{equation}
calculated according to published reconstruction resolutions~\cite{icecube_energy_reco, Aartsen_2017_monood}.
Angular error in reconstruction is nominalized with a Kent Distribution~\cite{10.2307/2984712} over azimuths to extract the zenith error.
The width of the Kent distribution, $\theta_{z}^{err}$, is energy-dependent according to the 50\% angular error presented in Ref~\cite{Aartsen_2017_monood}, so we solve for the parameter \(\kappa\) using
\begin{equation}
    -\int_{1}^{\cos\theta_{z}^{err}(E_{\nu}^{true})} \frac{\kappa}{2\sinh\kappa} e^{\kappa \cos\theta} d\cos\theta = 0.50
\end{equation}
at each analysis bin.

\begin{figure}
    \centering
    \includegraphics[width=0.95\linewidth]{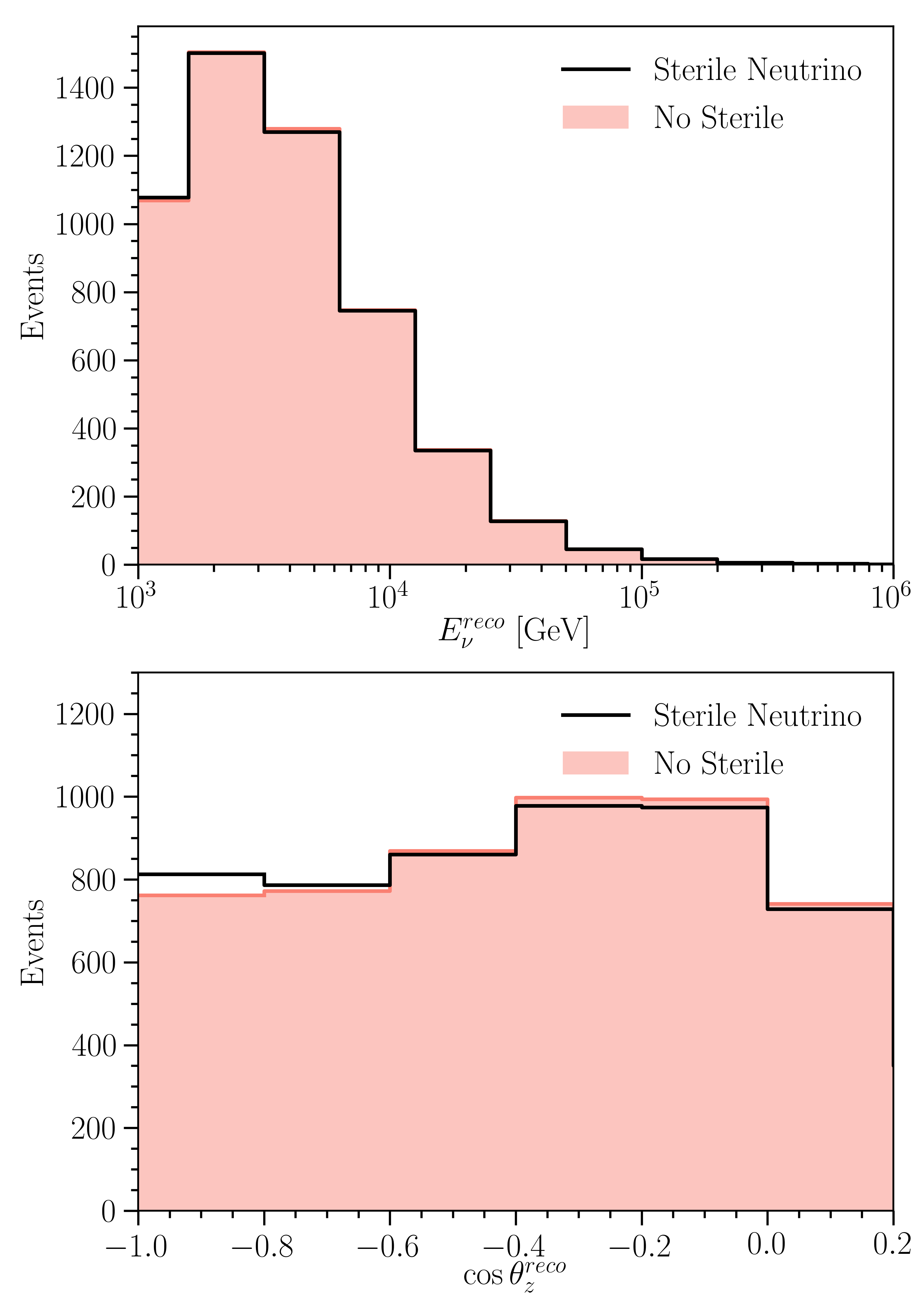}
    \caption{Expected number of cascades for ten years of livetime using the Hillas Gaisser H3a cosmic-ray flux model, SYBILL 2.3c interaction model, and the event selection described in Ref~\cite{2018PhDT17N} for a three-neutrino model (salmon) and a 3+1 sterile neutrino model (black line) with $\sin^{2}(2\theta_{24})=0.1$, $\sin^{2}(2\theta_{34})=0.2$, and $\Delta m_{41}^{2}=4.5\text{ eV}^{2}$. Number of events is summed over zenith angles (top) and energy (bottom) bins.  Note that the oscillation signature is a correlated function of both variables, so appears very indistinctly in these projections.}\label{fig:flat}
\end{figure}

The expected number of events for ten years of livetime is shown on the bottom of Figure~\ref{fig:eventrate}, and one-dimensional histograms of the number of events are shown in Figure~\ref{fig:flat}.

\subsection{Track Rates}
In order to calculate the expected track event rate in IceCube, we use a Monte Carlo set published as part of a previous 1-year search for sterile neutrinos~\cite{PhysRevLett.117.071801}. 
We use the same energy and cosine-zenith binning used in Subsection~\ref{sub:cascade}, and similarly scale the data to ten years of livetime. 
The expected number of tracks for ten years of livetime is shown in Figure~\ref{fig:eventrate}, top panel. 

\section{\label{sec:syst} Systematic Uncertainties}

A detailed treatment of IceCube's sources of systematic uncertainties would be prohibitively complex and require proprietary IceCube tools, so a truly rigorous sensitivity calculation for each channel can only be provided by the IceCube collaboration. Nevertheless, to estimate the expected impact of such effects we use publicly available data from Ref.~\cite{Aartsen_2020_prd} to apply a simplified treatment of the expected scale of systematic uncertainties. 
Dominant sources of systematic uncertainty are expected to derive from the shape and normalization of the atmospheric and astrophysical neutrino fluxes and the properties of the South Pole ice as in Ref.~\cite{Aartsen_2020_prd}. Other sources of uncertainty, such as the efficiencies of IceCube's DOMs, neutrino and anti-neutrino interaction cross sections, were not used in this preliminary analysis since they are subleading effects.

Absorption and scattering of light in the ice are treated using the effective gradient approach developed in Ref.~\cite{Aartsen_2019_snow} and used by Ref.~\cite{Aartsen_2020}. 
Uncertainties in the depth-dependence of the absorption and scattering of South Pole ice leads to uncertainties in energy reconstruction, and therefore an uncertainty in the energy-spectrum of expected event rates. 

The one-sigma deviations to the cosmic-ray flux are considered as in Ref~\cite{Aartsen_2020_prd}. 
These deviations calculate the expected one-sigma shifts in the expected atmospheric neutrino rates. 
Similarly, we perturb the slope of the astrophysical neutrino flux to determine variances in expected astrophysical neutrino rates. Per-bin uncertainties are then summed in quadrature to calculate a net systematic uncertainty.

Overall normalization of fluxes is treated as a nuisance parameter and allowed to float freely, such that we are studying energy and zenith shape and flavor ratio effects only, and not the absolute neutrino rate. We fit to the normalization before calculating the log likelihood at each physics point.  As will be described below, this simplified prescription has been tested by regenerating IceCube's sensitivity to $\theta_{24}$ via $\nu_\mu$ disappearance, and a similar median sensitivity to the published IceCube analysis is obtained (shown in Figure~\ref{fig:meowssense}).  Although both imperfect and incomplete, we believe that this prescription captures the majority of the important effects of the relevant systematic uncertainties for present purposes.


\section{\label{sec:results} Predicted Sensitivities}

A binned-likelihood approach is used in calculating the log-likelihood for the expected numbers of events for each set of physical parameters. The test statistic at each point in parameter space is calculated according to,
\begin{equation}
\text{TS} = -2\Delta LLH = -2\left(\ln L - \ln L_{max}\right),
\end{equation}
after removing the overall normalization effect by fitting the no-sterile-neutrinos flux to the parameter point of interest and adjusting the hypothesis normalization accordingly. We have performed likelihood based analyses in three samples, tracks only, cascades only, and tracks and cascades combined.

\subsection{\label{sec:tracksens} Tracks-only sensitivity to $\nu_\mu$ disappearance}

\begin{figure}
    \centering
    \includegraphics[width=0.99\linewidth]{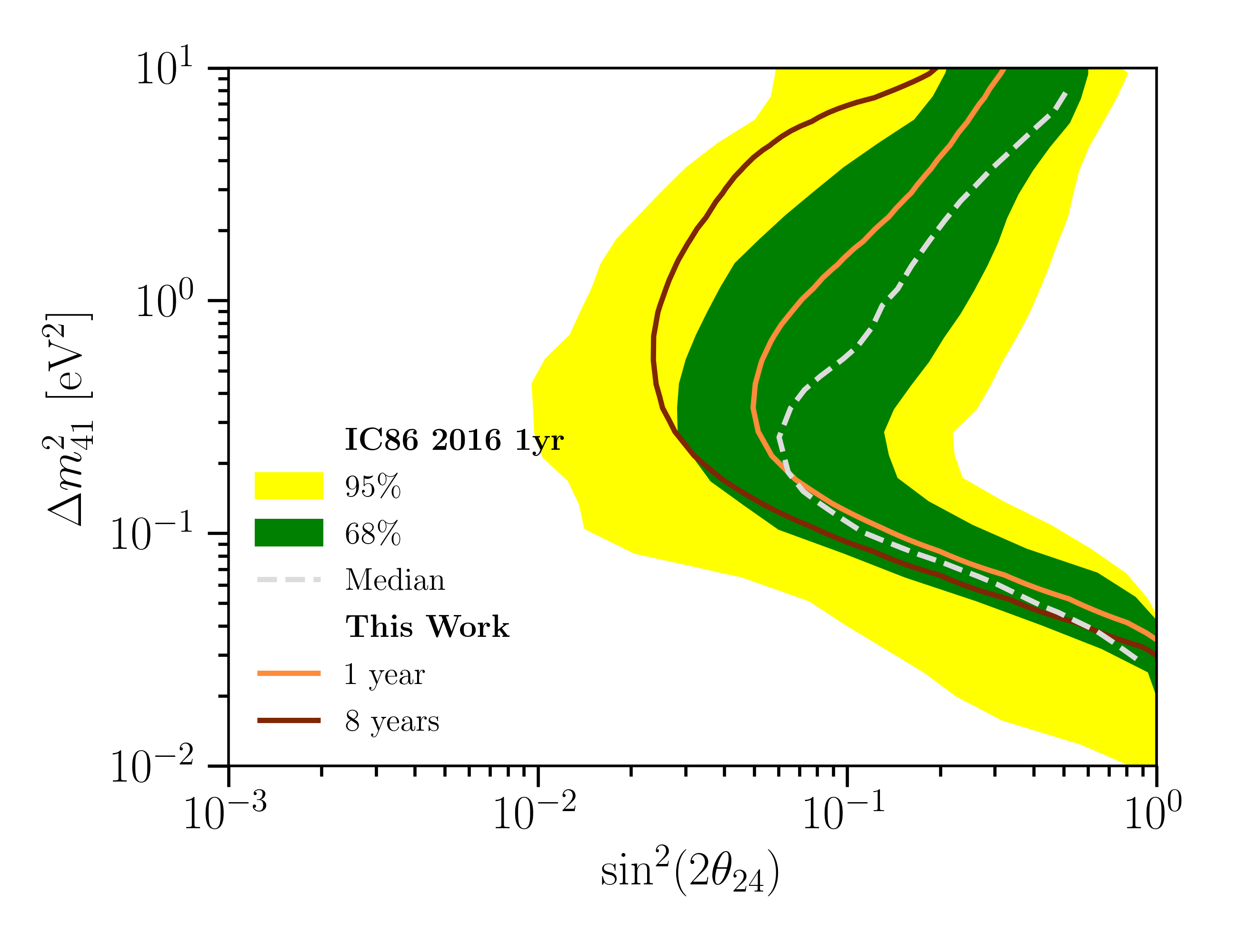}
    \caption{2 DOF, 90\% C.L. sensitivity to $\abs{U_{24}}^{2}$ and $\Delta m_{41}^{2}$ with $\theta_{34}=0.0$. These results closely reproduce those of Ref.~\cite{Aartsen_2020_prd}.}\label{fig:meowssense}
\end{figure}

We first perform likelihood analysis for IceCube's track sample: calculating the sensitivity to $\theta_{24}$ and $\Delta m_{41}^{2}$, using the procedure described in Section~\ref{sec:flux} to predict the expected number of tracks in only eight years of livetime. These results are shown in Figure~\ref{fig:meowssense}, which accurately reproduce the sensitivities presented in Ref.~\cite{PhysRevLett.117.071801}.  We have chosen this point of comparison rather than the more recent results of Ref.~\cite{Aartsen_2020} as the updated event selection there improves efficiencies at low energy, while the data release required to make these studies is only available at present for the earlier, one-year analysis. Approximate agreement of the median sensitivity, well within the bounds of expected fluctuations, validates our simplified analysis methodology as capturing the essential elements needed for a robust sensitivity estimate.  For completeness we also present an eight-year projection.  


\subsection{\label{sec:cascadesens} Cascades-only sensitivity to $\nu_{\tau}$ appearance}

\begin{figure}
    \centering
    \includegraphics[width=0.95\linewidth]{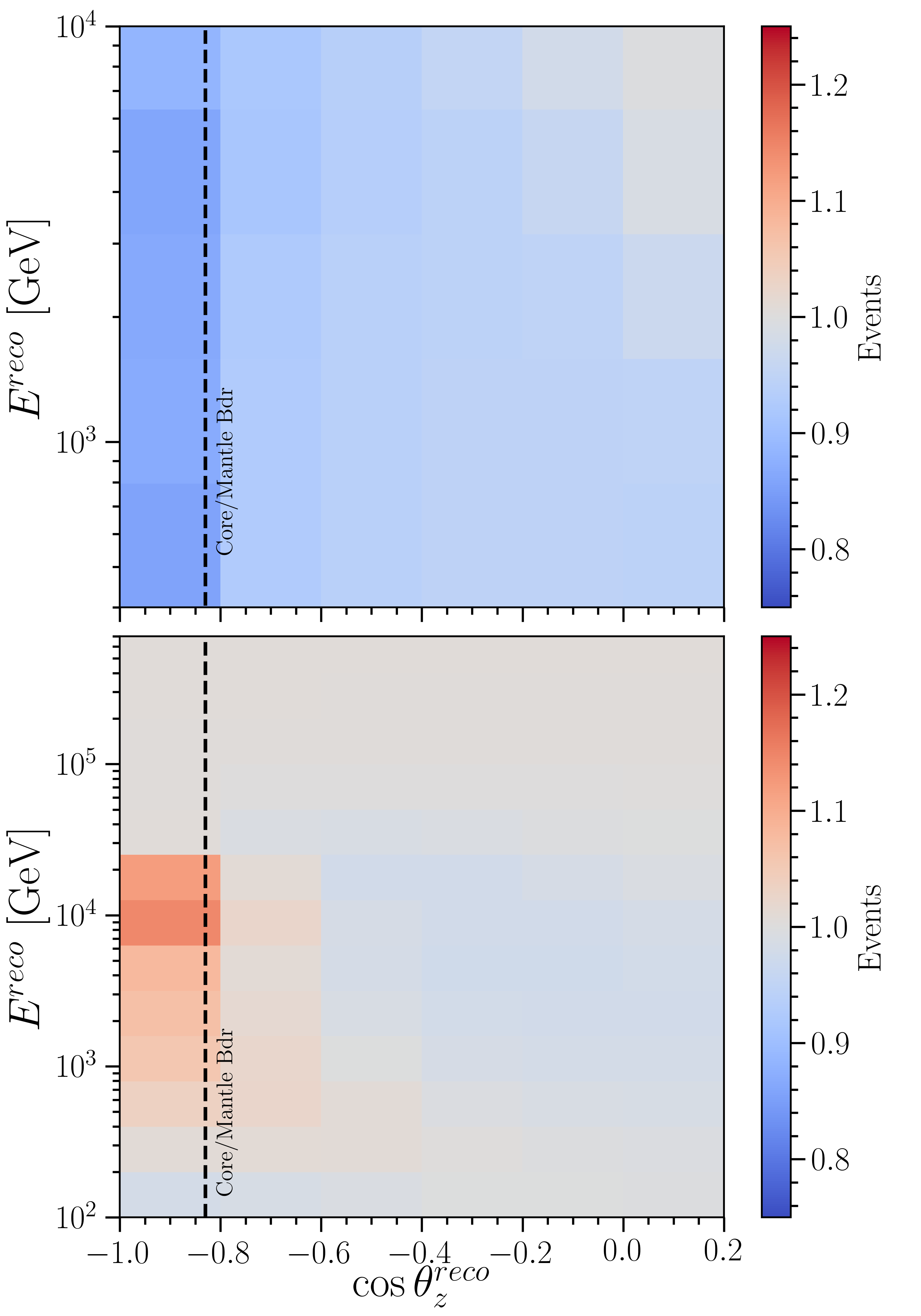}
    \caption{Ratio of expected tracks (top) and cascades (bottom) for a sterile neutrino model with $\sin^{2}(2\theta_{24})=0.1$, $\sin^{2}(2\theta_{34})=0.2$, and $\Delta m_{41}^{2}=4.5\text{ eV}^{2}$ and the standard three-neutrino model. Fluxes are calculated using the Hillas-Gaisser H3a cosmic ray flux model and the SYBILL 2.3c interaction model. A broad disappearance is expected in up-going tracks coincident with an appearance of up-going cascades.}
    \label{fig:appear}
\end{figure}

Signatures of $\nu_{\tau}$ appearance require nonzero values for all of  $\Delta m_{41}^{2}$, $\theta_{24}$, and $\theta_{34}$.  An example of a point with a non-trivial appearance signature that is consistent with existing experimental limits is shown in Figure~\ref{fig:appear}. This signature in reconstructed space is calculated by fixing $\theta_{24}$ and $\Delta m^2_{41}$ at their best-fit points from IceCube's $\nu_\mu$ disappearance searches, and fixing $\sin^{2}(2\theta_{34})=0.2$, comfortably consistent with current bounds, which are around $\sin^{2}(2\theta_{34})\lesssim 0.6$~\cite{Aartsen_2017_dc, Adamson_2011}.

Since all three of the above parameters must be non-zero to observe $\nu_\tau$ appearance, sensitivities should be expressed in three dimensional spaces (or four, if $\theta_{14}$ is also included). However, to facilitate  presentation of results on 2D plots in this work we have primarily opted to present two dimensional sensitivities under specific and experimentally motivated assumptions on the third parameter.

\begin{figure}
    \centering
    \includegraphics[width=0.99\linewidth]{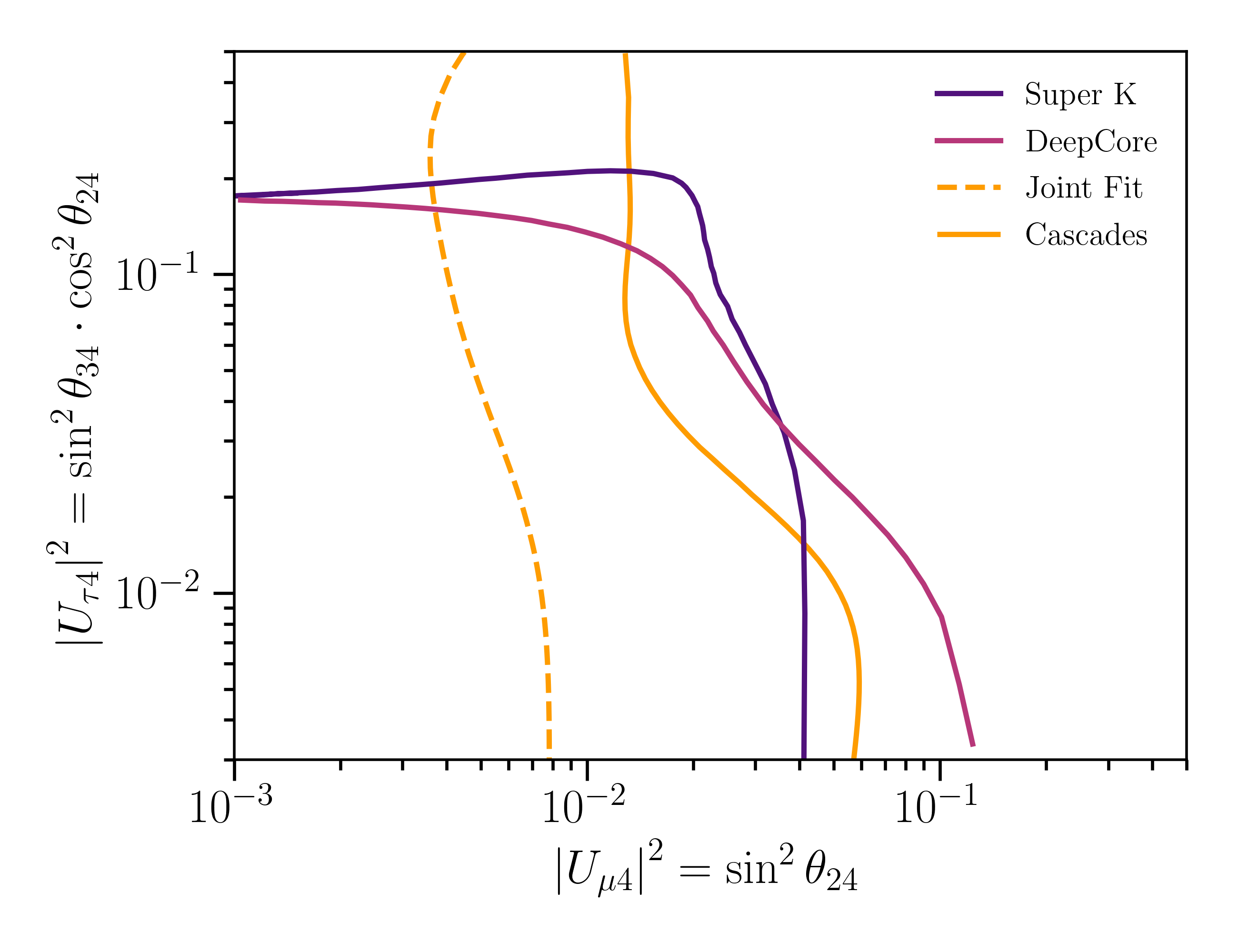}
    \caption{The 2 DOF, 90\% C.L. sensitivity to the $\theta_{24}$ and $\theta_{34}$ neutrino mixing parameters from this work with $\Delta m_{41}^{2}=1$ eV$^{2}$ for this work, IceCube's DeepCore~\cite{Aartsen_2017_dc}, and Super-Kamiokande~\cite{PhysRevD.91.052019}. The sensitivity through cascades is shown in the solid contour, and the joint track-cascade contour is dashed.}\label{fig:jointplot}
\end{figure}

\begin{figure}
    \centering
    \includegraphics[width=0.99\linewidth]{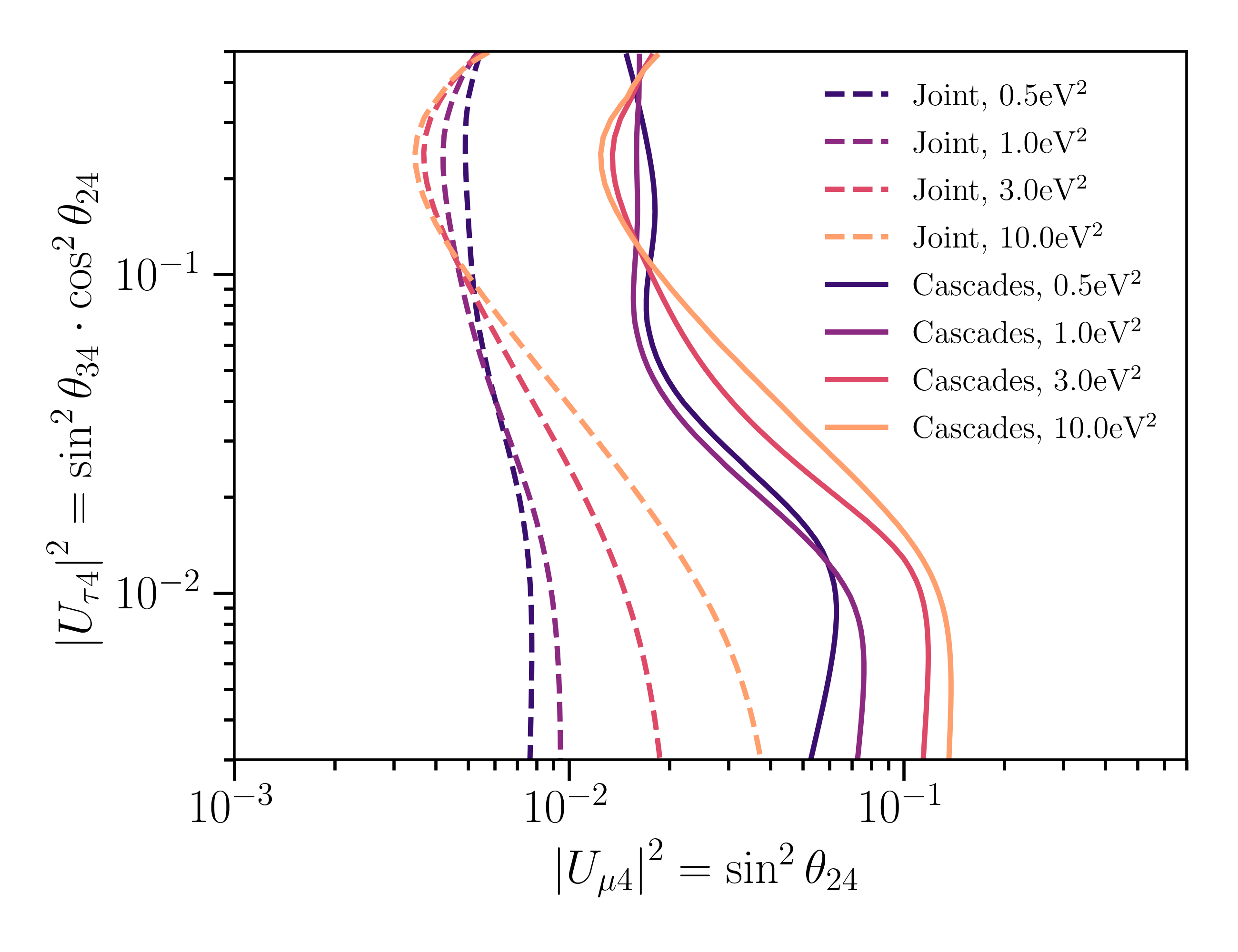}
    \caption{Cross-sections of the 3 DOF, 90\% C.L. sensivitivy surface to $\theta_{24}$,  $\theta_{34}$, and $\Delta m_{41}^{2}$. The sensitivity through cascades is shown in the solid contour, and the joint track-cascade contour is dashed.}\label{fig:manyplot}
\end{figure}

Using the methods described in Section~\ref{sec:flux}, we calculate expected cascade rates in IceCube at combinations of $\theta_{24}$, $\theta_{34}$, and $\Delta m_{41}^{2}$. 
The effects of $\theta_{14}$ are marginal unless large mixing angles are reached, and so for this part of the analysis it was kept to zero. 
The matter effects on these oscillations are similarly only marginally affected by the CP-violating phases~\cite{Aartsen_2020_prd}, and so they are fixed to zero.
The results of the sensitivity scan over cascade events only are shown in the solid line of Figure~\ref{fig:jointplot}, with sensitivities from other experiments overlaid, at the conventional benchmark point of $\Delta m^2_{41}=1\text{ eV}^2$; sensitivities at other values of $\Delta m^{2}_{41}$ are shown in the solid lines of Figure~\ref{fig:manyplot}. We see that with cascades alone we expect a sensitivity competitive with other leading sensitivities from Super-Kamiokande~\cite{PhysRevD.91.052019} and IceCube's DeepCore~\cite{Aartsen_2017_dc}. Sensitivities are the most competitive for points in phase space where both $\theta_{24}$ and $\theta_{34}$ are large; here, the transition probability $P(\nu_\mu\to\nu_{\tau})$ is maximized. 

Meanwhile, in regions where $\abs{U_{\tau 4}}^{2}$ is small, $\nu_{\mu}$ disappearance is most significant in a signal similar to Refs.~\cite{Aartsen_2020, Aartsen_2020_prd}, but as $\nu_{\mu}$ cascades.
A small increase to $\abs{U_{\tau 4}}^{2}$ can then lead to competing $\nu_{\tau}$ appearance and $\nu_{\mu}$ disappearance, and so for small values of $\Delta m_{41}^{2}$, this leads to a reduction of sensitivity. 
Since $\nu_{\mu}$ events overwhelmingly lead to cascades while $\nu_{\tau}$ often cause tracks, at higher $\abs{U_{\mu 4}}^{2}$ the $\nu_{\tau}$ appearance begins to dominate and sensitivity improves.
Finally, since tau appearance follows a $\nu_{\mu}\to\nu_{s}\to\nu_{\tau}$ appearance channel, a non-zero $\abs{U_{\mu 4}}^{2}$ is needed to for any sensitivity; this causes a lower bound on the $\abs{U_{\mu 4}}^{2}$ sensitivity. 

\subsection{\label{sec:joint} Joint Sensitivity for $\nu_\mu$ disappearance and $\nu_\tau$ appearance} 

By performing a joint sensitivity using both cascade and track-like events, we are able to significantly improve the sensitivity, by exploring a flavor ratio rather than a pure shape effect. Track-like events will provide a method to fit to the overall flux normalization and further constrain sensitivities. 
Specifically, the process described in in Subsection~\ref{sec:cascadesens} is performed for track events, and the fit event-number normalization is then used in calculating the log likelihood in the cascade channel. The combined likelihood for both is then used in determining sensitivity contours. 
These results are shown in Figure~\ref{fig:jointplot}. A significant sensitivity enhancement relative to either tracks or cascades alone is obtained.

\begin{figure}
    \centering
    \includegraphics[width=0.99\linewidth]{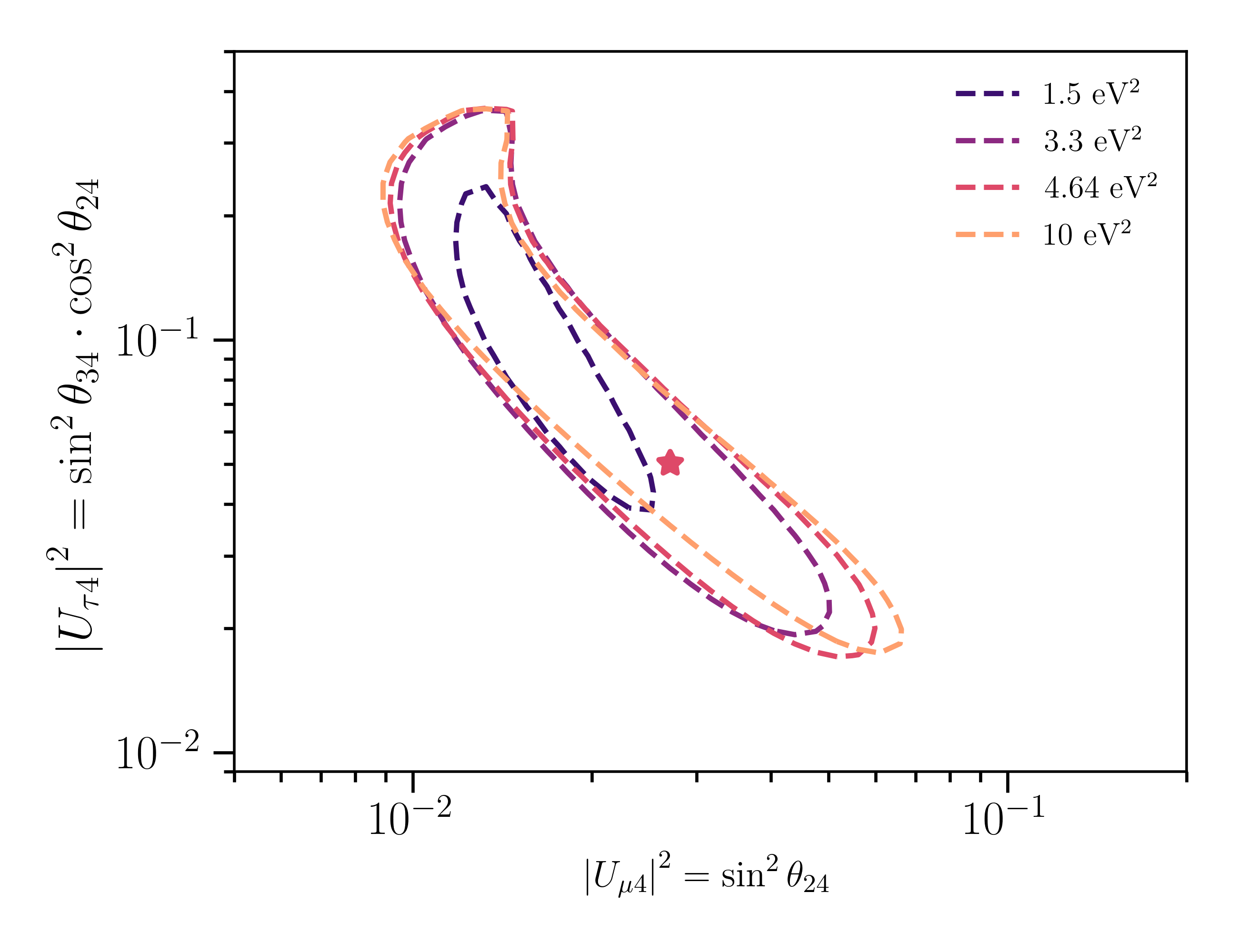}
    \caption{3 DOF, 90\% C.L. sensitivity to the $\abs{U_{\mu 4}}^{2}$ and $\abs{U_{\tau 4}}^{2}$ PMNS matrix elements, at various values of $\Delta m_{41}^{2}$, for this work using a joint track-cascade likelihood assuming a sterile neutrino with $\sin^{2}(2\theta_{24})=0.1$, $\sin^{2}(2\theta_{34})=0.2$, and $\Delta m_{41}^{2}=4.64\text{ eV}^{2}$.}\label{fig:jointsterile}
\end{figure}

In addition to calculating sensitivity, we examine the results that may be expected in the presence of a sterile neutrino with non-zero $\theta_{24}$ and $\theta_{34}$.  In Figure~\ref{fig:jointsterile} we show the result obtained by injecting a signal with $\sin^2(2\theta_{24})=0.1$, $\sin^2(2\theta_{34})=0.2$ and $\Delta m^2_{41}=4.64\text{ eV}^2$ and fitting over values of the mixing parameters; this mass squared splitting was chosen out of computational convenience as it lines up with a point at which fluxes were calculated.
We include four slices through the space in $\Delta m_{41}^2$ at several benchmark points, and provide contours at 90\% CL calculated using $\chi^2$ assuming that the test statistic, TS, satisfies Wilk's theorem and is distributed with a $\chi^{2}$ distribution with thresholds with thresholds consistent with three degrees of freedom.  A signature of this form, which is consistent with present constraints would be potentially discoverable in a joint tracks and cascades analysis at IceCube.

\subsection{\label{sec:probebest} Joint Sensitivity for $\nu_\mu$ and disappearance generic cascade appearance}

\begin{figure}
    \centering
    \includegraphics[width=0.95\linewidth]{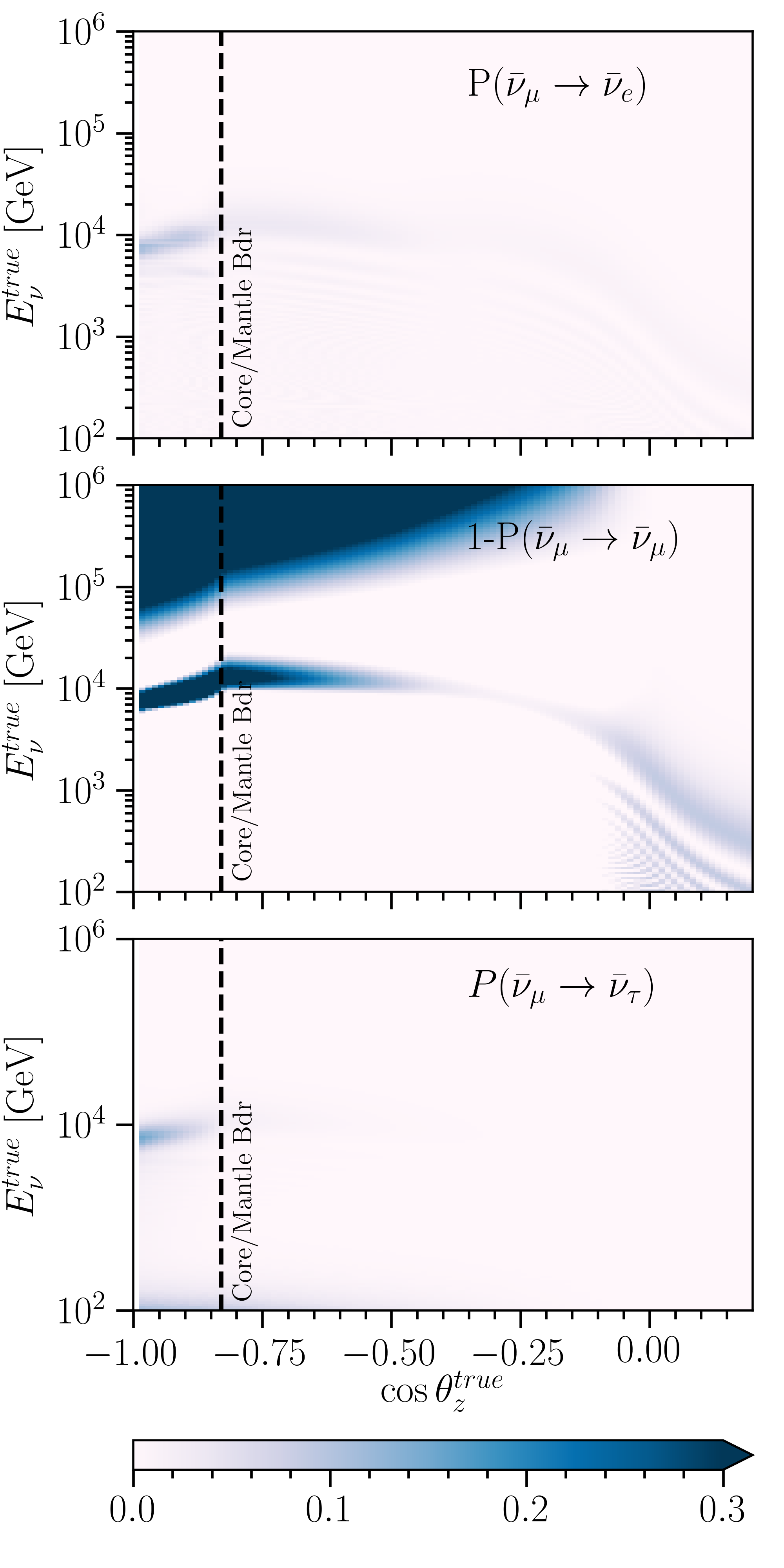}
    \caption{Transition probabilities $P(\bar{\nu}_{\mu}\to\bar{\nu}_{\alpha})$ for $\bar{\nu}_e$ (top), $\bar{\nu}_{\mu}$ (middle), and $\bar{\nu}_{\tau}$ (bottom) for a sterile neutrino flux with \(\sin^{2}(2\theta_{14})=0.43\), \(\sin^{2}(2\theta_{24})=0.1\), \(\sin^{2}(2\theta_{34})=0.01\),  and \(\Delta m^{2}_{41}=3.3\text{ eV}^{2}\). A dashed black line is used to denote the outer core-mantle boundary.}\label{fig:bestosc}
\end{figure} 

Cascade appearance may be introduced not only from $\nu_\tau$ appearance, but also from $\nu_e$ appearance.  A non-zero value for $\theta_{14}$ is motivated in particular by recent results from the BEST experiment, which motivates us to consider whether IceCube fitting both cascade and track channels has sensitivity to values of $\theta_{14}$ consistent with such a $\nu_e$ disappearance effect.  IceCube can of course not rule out the BEST anomaly alone, since a scenario with $\theta_{24}=0$ will generate no substantial appearance signatures in IceCube for any value of $\theta_{14}$.  But in principle it may be able confirm the BEST anomaly, given sizeable enough values for both $\theta_{24}$ and $\theta_{14}$. In such a model with non-zero $\theta_{14}$, $\theta_{24}$, and $\theta_{34}$, resonant oscillations lead to appearances in both the $\nu_{\tau}$ and $\nu_{e}$ channels shown in Figure~\ref{fig:bestosc}, wherein the BEST best-fit values were used for $\theta_{14}$ and $\Delta m_{41}^{2}$ 

\begin{figure}
    \centering
    \includegraphics[width=0.99\linewidth]{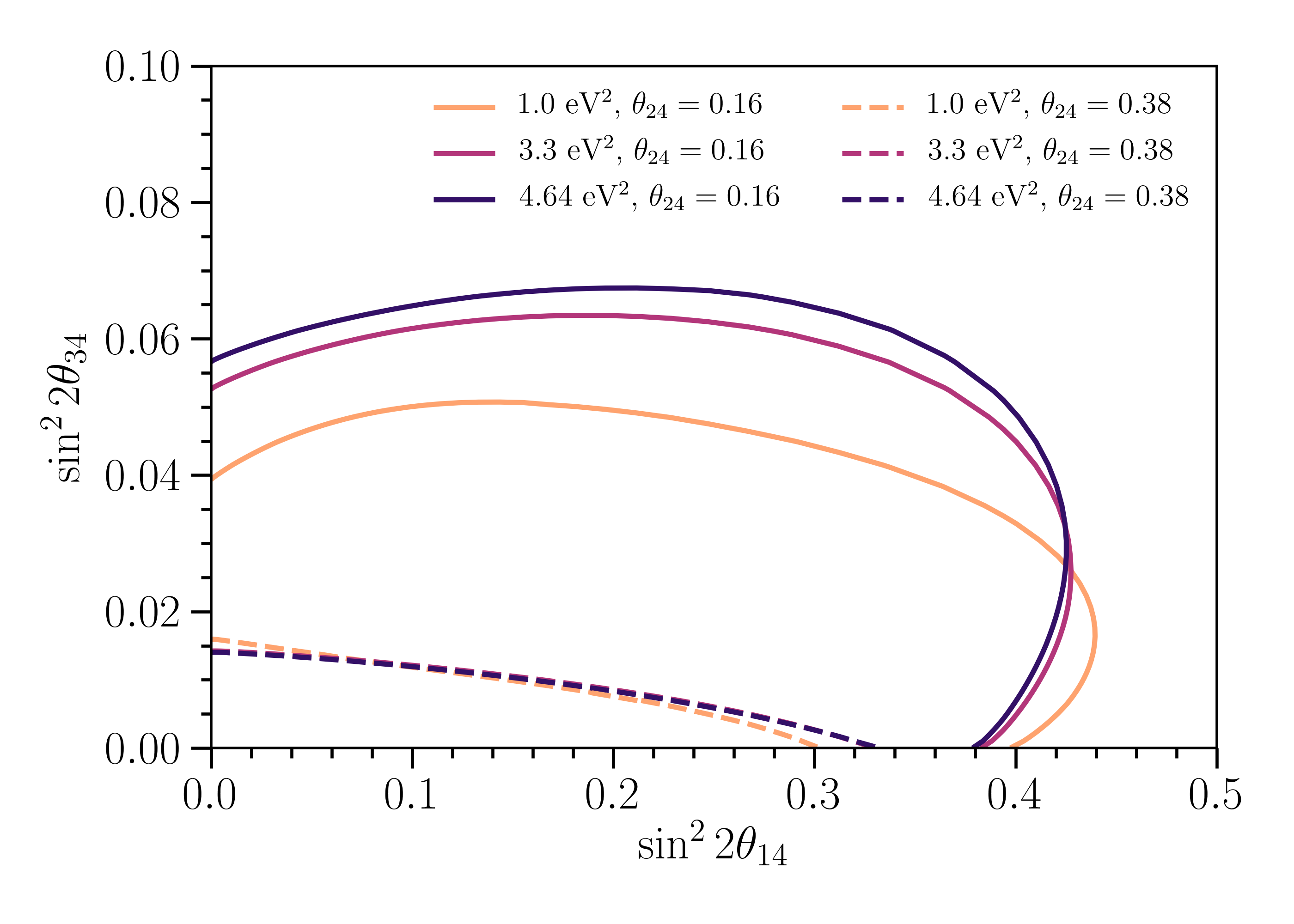}
    \caption{The 2 DOF, 90\% C.L. sensitivity contours for $\sin^{2}\theta_{14}$ and $\sin^{2}\theta_{34}$, using a joint track-cascade likelihood, for a 3+1 sterile neutrino model with various values of $\Delta m_{41}^{2}$ and $\theta_{24}$, and $\theta_{14}=0.0$.}\label{fig:bestfirst}
\end{figure}

To assess sensitivity to this effect in IceCube, scans over $\theta_{14}$ and $\theta_{34}$ were performed at multiple values of $\Delta m_{41}^{2}$ and $\theta_{24}$. 
In Figure~\ref{fig:bestfirst} we show IceCube's sensitivity to a 3+1 sterile neutrino model with $\theta_{14}=0.0$. These contours represent the median expected 90\% confidence level that could be drawn if no sterile neutrino were present, given assumptions on the non-plotted parameters shown in the caption.  The two choices of assumptions made on the non-fitted parameters correspond to $\theta_{24}=0.1609$ (the $\nu_\mu$ disappearance best fit point from Ref.~\cite{Aartsen_2020}) or $\theta_{24}=0.3826$ (a value within the 90\% results contour of Ref.~\cite{Aartsen_2020}), and $\Delta m^2_{41}=1\text{ eV}^2$ (a standard benchmark point in the field), $\Delta m^2_{41}=3.3\text{ eV}^2$ (the BEST best-fit point), and $\Delta m^2_{41}=4.64\text{ eV}^2$ (close to the IceCube $\nu_\mu$ disappearance best fit point at $4.5\text{ eV}^2$).  It is observed that IceCube has significant sensitivity in this high dimensional parameter space for many values of the mixing parameters consistent with the present BEST and IceCube results, assuming a non-zero value of $\theta_{24}$ consistent with IceCube's existing preferred regions from $\nu_\mu$ disappearance measurements.

\begin{figure}
    \centering
    \includegraphics[width=0.99\linewidth]{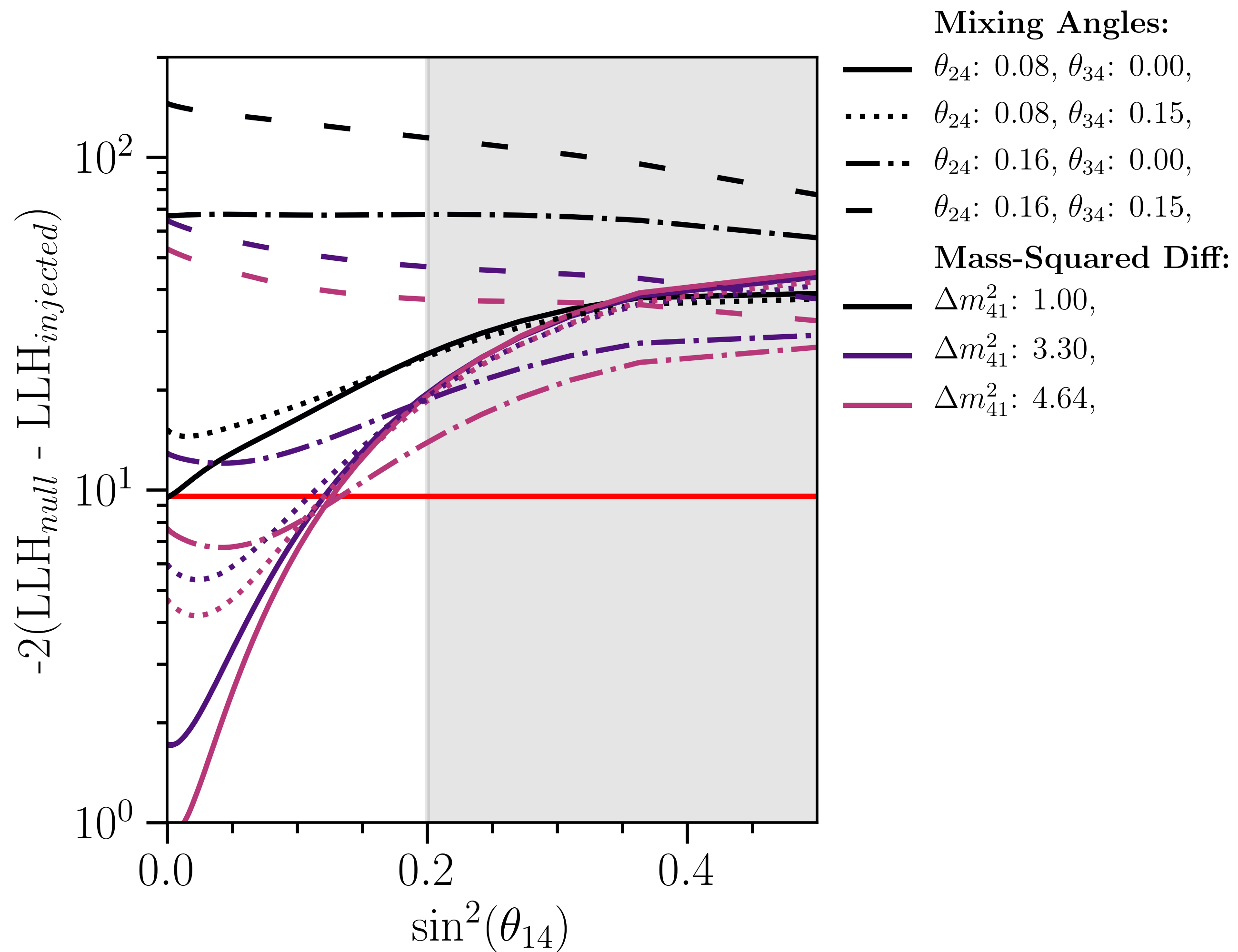}
    \caption{The test statistics values for various different injected 3+1 sterile neutrino models, using a joint track-cascade likelihood, compared to a three-neutrino hypothesis. The red line represents a 4 DOF 95\% CL sensitivity threshold; the shaded region represents the 95\% confidence level bounds from the BEST best fit.}\label{fig:bestsecond}
\end{figure}

A more intuitive picture of IceCube's capability to confirm the BEST anomaly as being sterile-neutrino related, given values of other mixing parameters consistent with IceCube and world data, is shown in Fig.~\ref{fig:bestsecond}. Here, likelihoods are calculated according to injected sterile neutrino parameters and assuming a three-neutrino model; the resulting test statistics are shown. For all considered combinations, IceCube is seen to be capable of discriminating a BEST-like sterile neutrino flux from a three-neutrino model at the 95\% confidence level. Thus, IceCube appears to have capability to confirm the best anomaly at least 95\% confidence, given suitable values of the other mixing parameters, within existing constraints and uncertainties.

\section{\label{sec:conclusion} Conclusions}

We have considered IceCube's sensitivity to sterile neutrinos through the cascade appearance channel. Both $\nu_\tau$ and $\nu_e$ appearance signatures are in principle observable by IceCube for $\theta_{14}$ and $\theta_{34}$ values within existing constraints, $\theta_{24}$ around IceCube's present preferred values from $\nu_\mu$ disappearance, and many possible values $\Delta m^2_{41}$. 

We find that IceCube's sensitivity in the joint $U_{\tau 4}$, $U_{\mu 4}$ space that has been explored by previous analyses at SuperKamiokande and IceCube will be enhanced significantly at the benchmark point of $\Delta m^2=1\text{ eV}^2$ by a joint fit to both track and cascade samples.  Strong sensitivity is also obtained for other mass points, under the standard mixing assumption of $\theta_{14}=0$.  Cascade signatures that may accompany tentative but weak hints of $\nu_\mu$ disappearance for $\Delta m^2_{41}\sim4.5\text{ eV}^2$ and $\sin^2(\theta_{24})\sim 0.1$ are discoverable at IceCube with values $\theta_{34}$ that remain consistent with world data, strongly motivating investigation of $\nu_\tau$ appearance via cascades in parallel with the established IceCube searches for $\nu_\mu$ disappearance using tracks.

We have also explored the effect of introducing non-trivial $\nu_e$ appearance, consistent with the BEST and gallium anomalies, via non-zero $\theta_{14}$.  IceCube cannot rule out the BEST or gallium preferred regions in $\theta_{14}$ alone, since sensitivity of IceCube to this parameter requires non-zero $\theta_{24}$. For modest values of $\theta_{24}$ at either the IceCube best fit point in $\nu_\mu$ disappearance or at a point near the 90\% CL upper limit in this channel, however, values of $\theta_{14}$ and $\Delta m^2_{41}$ around the best fit point can be probed at better than 90\% confidence level.  

We conclude that our joint analysis of track and cascade topologies at IceCube can contribute to the ongoing worldwide project of understanding short baseline anomalies in both $\nu_e$ appearance and disappearance channels.  The IceCube data set, probing both $\nu_\mu$ disappearance and $\nu_e$ and $\nu_\tau$ appearance near the matter resonance for core crossing neutrinos, provides unique and powerful insights into possible mixing of heavier neutrino mass states with the $\nu_\tau$ flavor, as well as offering sensitivity to $\nu_e$ appearance in experimentally relevant parts of parameter space associated with the BEST and gallium anomalies.

\section{Acknowledgments}
BS and BJPJ are supported by the National Science Foundation under award number 1913607. CAA is supported by the Faculty of Arts and Sciences of Harvard University, and the Alfred P. Sloan Foundation. JC and AD are supported by the National Science Foundation under award number 1912764.

\bibliographystyle{aipauth4-1}
\bibliography{main}
\clearpage

\end{document}